\documentclass[prl,twocolumn,floatfix,superscriptaddress,showpacs]{revtex4}
\usepackage{graphicx,dcolumn,bm}

\begin{document}

\title{Polarization and relaxation of radon}

\def\groupUMF{\affiliation{FOCUS Center, University of Michigan Physics Department, 450 Church St., Ann Arbor 48109-1040, USA}}
\def\groupUM{\affiliation{University of Michigan Physics Department, 450 Church St., Ann Arbor 48109-1040, USA}}
\def\groupTRIUMF{\affiliation{TRIUMF, 4004 Westbrook Mall, Vancouver V6T 2A3, Canada}}
\def\groupSUNYSB{\affiliation{SUNY Stony Brook Department of Physics and Astronomy, Stony Brook 11794-3800, USA}}

\groupUMF
\groupUM
\groupTRIUMF
\groupSUNYSB

\author{E. R. Tardiff} \groupUMF
\author{J. A. Behr} \groupTRIUMF
\author{T. E. Chupp} \groupUMF
\author{K. Gulyuz} \groupSUNYSB
\author{R. S. Lefferts} \groupSUNYSB
\author{W. Lorenzon} \groupUM
\author{S. R. Nuss-Warren} \groupUMF
\author{M. R. Pearson} \groupTRIUMF
\author{N. Pietralla} \groupSUNYSB
\author{G. Rainovski} \groupSUNYSB
\author{J. F. Sell} \groupSUNYSB
\author{G. D. Sprouse} \groupSUNYSB

\date{\today}
 

\begin{abstract}
Investigations of the polarization and relaxation of $^{209}$Rn by spin exchange with laser optically pumped rubidium are reported.  On the order of one million atoms per shot were collected in coated and uncoated glass cells.  Gamma-ray anisotropies were measured as a signal of the alignment (second order moment of the polarization) resulting from the combination of polarization and quadrupole relaxation at the cell walls.  The temperature dependence over the range 130$^\circ$C to 220$^\circ$C shows the anisotropies increasing with increasing temperature as the ratio of the spin exchange polarization rate to the wall relaxation rate increases faster than the rubidium polarization decreases.  Polarization relaxation rates for coated and uncoated cells are presented.  In addition, improved limits on the multipole mixing ratios of some of the main gamma-ray transitions have been extracted.  These results are promising for electric dipole moment measurements of octupole-deformed $^{223}$Rn and other isotopes, provided sufficient quantities of the rare isotopes can be produced.
\end{abstract}

\pacs{32.10.Dk, 11.30.-j, 23.20.En, 23.20.Gq}
\maketitle


  An electric dipole moment (EDM) is a T-violating separation of charge along the spin of a system such as the neutron, an atom, or a molecule.  Assuming CPT symmetry, EDM measurements probe CP violation in the Standard Model and beyond.  An atomic EDM would arise as a result of the presence of CP-odd fundamental interactions that affect physics at the atomic scale.  Possible examples include the vacuum expectation value of the gluon field ($\theta_{QCD}$), fundamental quark and electron EDMs, CP-odd four-quark interactions \cite{Barr93}, and the Weinberg 3-gluon interaction \cite{Barr93,Wein89}.  The search for permanent EDMs has implications for many areas of beyond-the-Standard-Model physics.  The expanded particle spectrum introduced by supersymmetric theories increases the number of diagrams contributing CP-violating phases \cite{Abel01}.  The baryon asymmetry in the universe could be generated by a mechanism that requires CP violation beyond that contained within the Standard Model \cite{SMCPNN} that could be revealed by EDM measurements. 
  
 Isotopes featuring octupole deformation or vibrational strength, such as $^{223}$Rn and $^{225}$Ra, are expected to exhibit a large enhancement to their sensitivity to a CP-violating EDM \cite{Spev97,Enge00}.  Calculations estimate that the octupole deformation of $^{223}$Rn makes it approximately 400 times more sensitive to CP violation than $^{199}$Hg \cite{Spev97},  which currently sets the best limits on several mechanisms of CP violation.  With a limit on the $^{199}$Hg EDM of $|d(^{199}\textrm{Hg})|<2.10\times10^{-28} e\cdot\rm{cm}$ \cite{Roma01}, a measurement of the $^{223}$Rn EDM with a sensitivity of 10$^{-26} e\cdot\rm{cm}$ would significantly improve sensitivity to CP violation.  NMR techniques have proven effective in making sensitive EDM measurements of some of the lighter noble gases \cite{Rose01}, but they have never been applied to EDM measurements of small amounts of short-lived radon isotopes.  In anticipation of high $^{223}\textrm{Rn}$ production rates at TRIUMF's ISAC \cite{ISAC04} and future facilities, a $^{209}$Rn source was developed at the Stony Brook Francium Lab \cite{Sims96}.  Although the polarization of $^{209}$Rn and $^{223}$Rn has been demonstrated \cite{Kita88}, measurements had been made at only one cell temperature. Both $^{209}$Rn (\textbf{I}=5/2, T$_{1/2}$=28.5 min) and $^{223}$Rn (\textbf{I}=7/2, T$_{1/2}$=23.2 min), when polarized in a glass optical pumping cell, are expected to have similar quadrupole interactions with the cell walls.  Thus $^{209}$Rn provides an appropriate system for the study of polarization and relaxation processes. 
 
\begin{figure}[h]
\begin{center}
\includegraphics[height=95mm]{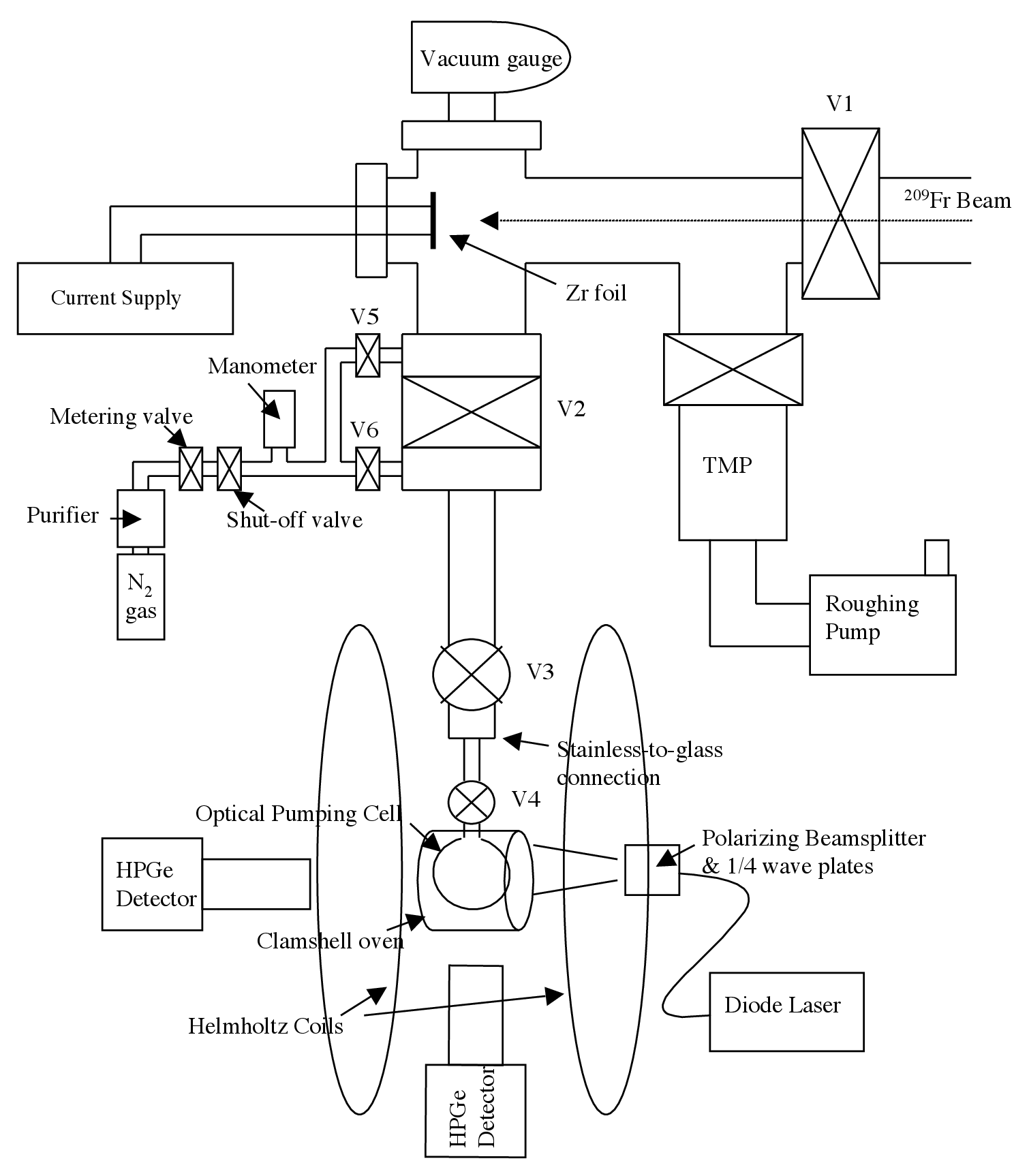}
\end{center}
\caption{A schematic diagram of the experimental apparatus.}
\label{fig:fig1}
\end{figure}
   
\begin{figure}[t]
\begin{center}
\includegraphics[height=95mm]{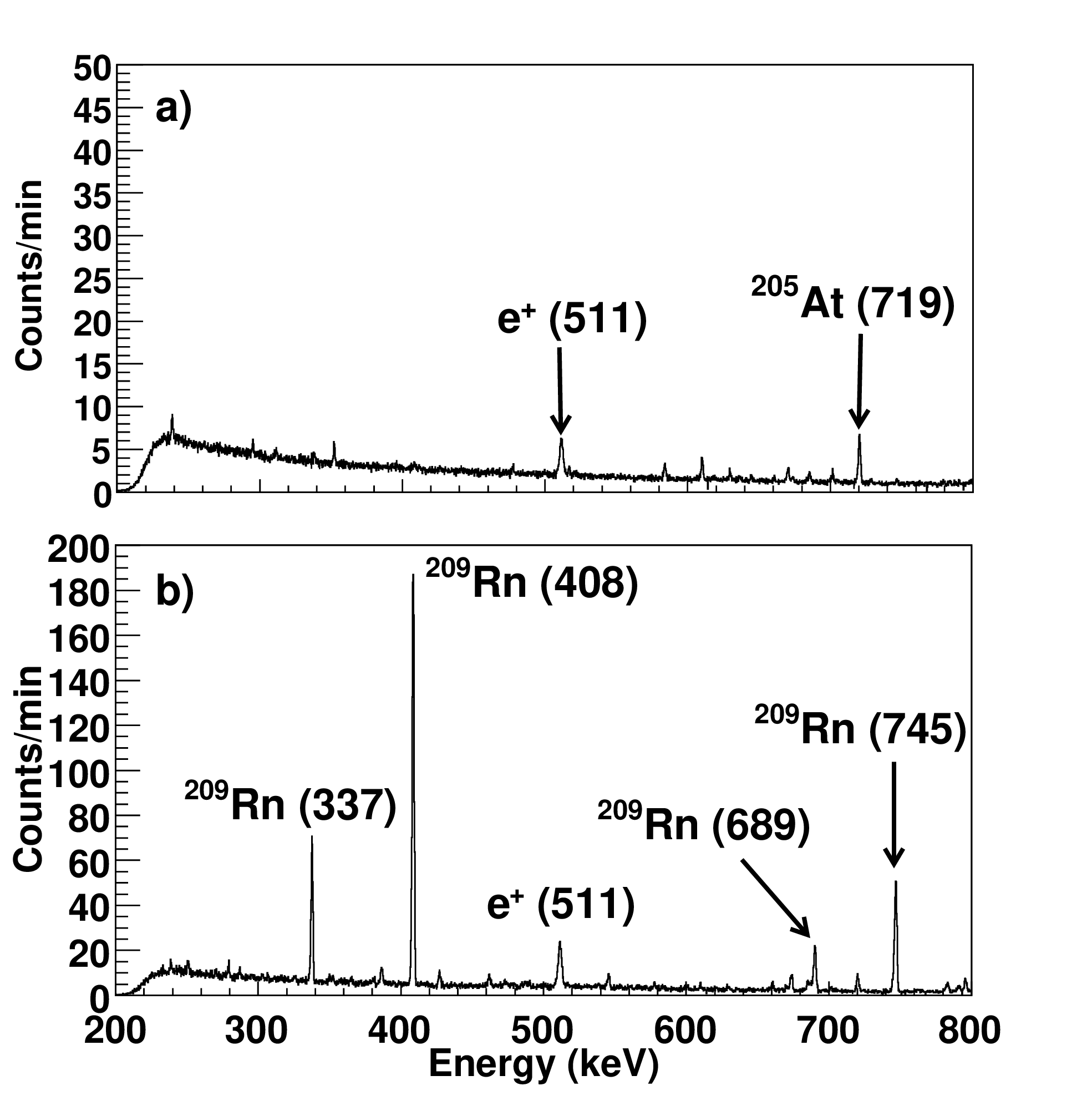}
\end{center}
\caption{Gamma-ray spectra from the 0$^\circ$ detector.  Spectrum a) was taken over about an hour during the implantation of the $^{209}$Fr beam in the Zr foil and b) was taken over the first ten minutes after transfer of the $^{209}$Rn to the optical pumping cell. Note the different vertical scales.}
\label{fig:fig2}
\end{figure}

  A beam of $^{16}$O incident on a gold target generated francium isotopes that were accelerated to 5 keV, electrostatically focussed, and implanted in a zirconium foil (see Fig. \ref{fig:fig1}).  Accelerating the $^{16}$O beam to around 91 MeV optimized the production, through the reaction $^{197}$Au($^{16}$O,4n), of $^{209}$Fr (T$_{1/2}=50.0$ s), which has an 11\% branching ratio for electron capture decay to $^{209}$Rn.  The francium was implanted for about two half-lives of $^{209}$Rn.  While there was some shielding between the foil and the detectors, the activity at the foil could still be monitored during implantation.  For example, the 719 keV line in the upper panel of Fig. \ref{fig:fig2} indicates $^{205}$At in the foil, the product of $^{209}$Fr $\alpha$-decay (89\% b.r.).  After implantation, the radon was transferred to the optical pumping cell, which had been preloaded with rubidium, by simultaneously immersing the cell in liquid nitrogen and heating the foil to about 1000$^\circ$C with the chamber isolated from the beamline and pumps.  At this temperature, radon rapidly diffused out of the foil \cite{Warn05}.  Valve V2 was then closed and about one atmosphere of nitrogen was added to the cell to act as a buffer gas, preventing radiation trapping.  The cell was then isolated from the rest of the vacuum system by shutting valve V4, and a two-piece glass oven was placed around it, allowing it to be heated to a desired temperature.  Up to about one million $^{209}$Rn atoms were transferred to the cell in each cycle.  In the first of two runs, discussed in Ref. \cite{Tard07} and designated Run 1 in this letter, data were taken for an uncoated cell at temperatures ranging from 150 to 200$^\circ$C.  In Run 2, data were obtained for an uncoated cell from 150 to 220$^\circ$C and for a cell coated with octadecyltrichlorosilane (OTS) \cite{Otei90} from 130 to 180$^\circ$C.
    
   The decay of polarized $^{209}$Rn results in the emission of $^{209}$At gamma rays with an angular distribution dependent on the alignment, i.e. the second moment of the nuclear sublevel population distribution resulting from temperature-dependent polarization and relaxation processes.  The cell was placed in a uniform magnetic field of about 10.5 gauss and illuminated by circularly polarized laser light from a Coherent diode laser system tuned to the Rb-D1 line.  This polarized the rubidium, and spin-exchange collisions \cite{Walk97} transferred the valence electron polarization to the $^{209}$Rn nuclei.  The resulting angular distribution of gamma rays was monitored by HPGe detectors: a Eurisys Clover at 0$^\circ$ (4 crystals, each with 25\% internal efficiency relative to NaI) and a 100\% Ortec detector at 90$^\circ$ in Run 1, and two 100\% Ortec detectors at both 0$^\circ$ and 90$^\circ$ (placed about two inches from the cell) in Run 2.  The data from the individual crystals in the Clover detector were summed in a way that allowed it to be viewed as one large crystal.  Data were also taken with the laser off in order to determine the unpolarized baseline rates at each detector and temperature.  Earlier studies of the wall interactions of the spin-3/2 isotope $^{131}$Xe \cite{Wu90} indicated reduced quadrupole relaxation in uncoated cells, which motivated the study of both coated and uncoated cells in these runs.

\begin{figure}[b]
\begin{center}
\includegraphics[width=87mm]{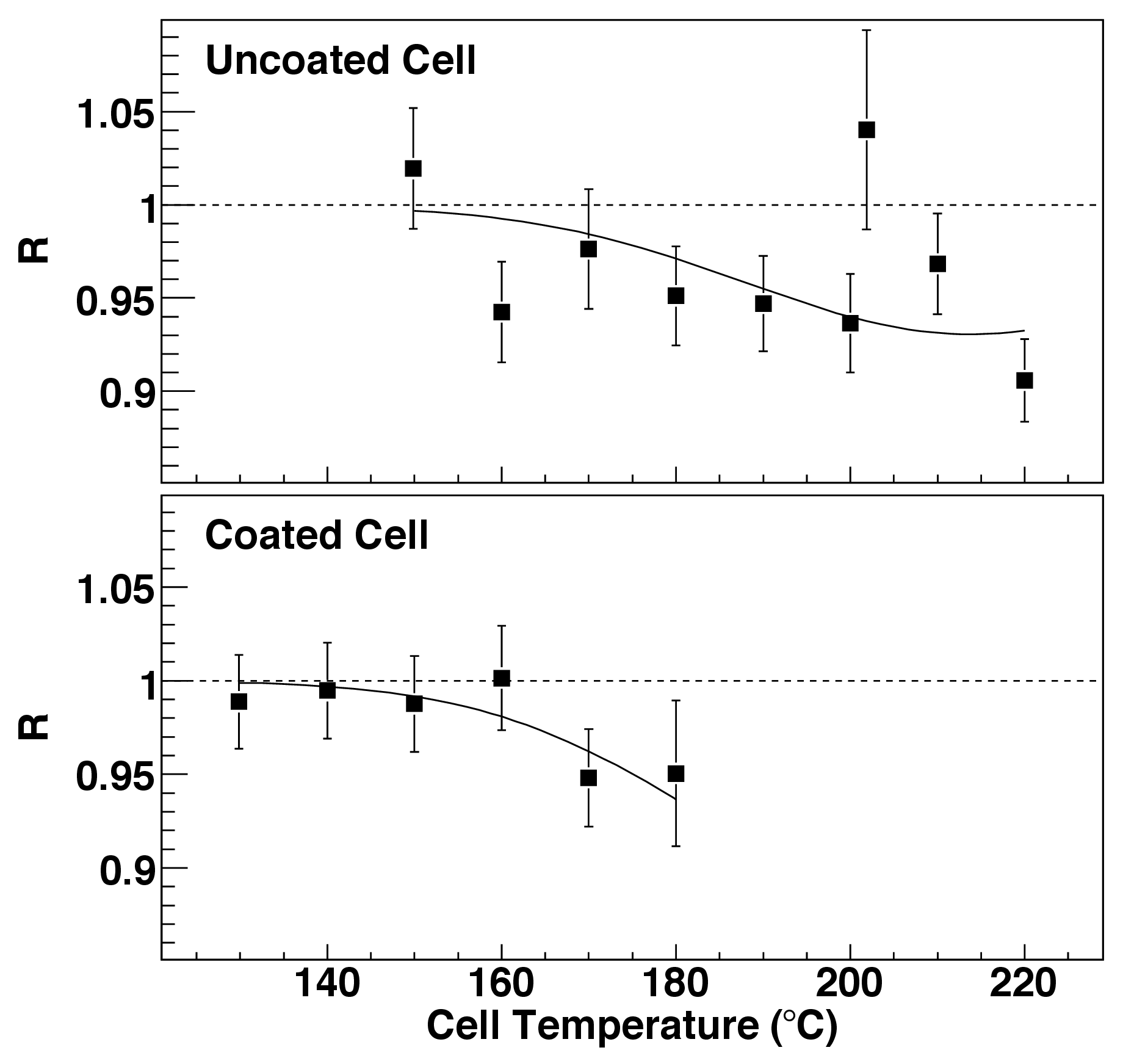}
\end{center}
\caption{Anisotropy data from the coated and uncoated cells for the 337 keV $^{209}$Rn gamma ray.  For the uncoated cell the average of Runs 1 and 2 is displayed.  R is the ratio defined in Eq. (4), and the solid curves are the fits, described in the text, from which a value for $\Gamma_2^\infty$ is obtained.}
\label{fig:fig3}
\end{figure}  

  In order to extract information on polarization and relaxation, an optical pumping/spin-exchange model was developed to find the radon steady-state nuclear sublevel populations \cite{Tard07b}.  The rubidium polarization is calculated by numerical integration of laser-light absorption in the cell with the Rb-Rb spin destruction rate from Ref. \cite{Bara98} and the rubidium number density given by \cite{Alco84}
\begin{equation}
[\textrm{Rb}]=10^{9.318-\frac{4040}{T}}/k_B T,
\end{equation}  
following the method detailed in Ref. \cite{Wags89}.  At low temperatures, the rubidium is almost completely polarized as optical pumping overcomes all spin destruction processes.  At higher temperatures, increased spin destruction with increased [Rb] leads to a decrease in the polarization due to limited laser power.  The resulting radon polarization results from the interplay of processes whose effects can be parameterized by three main rates: the rubidium-radon spin-exchange rate $\gamma_{SE}$, the radon dipole relaxation rate $\Gamma_1$, and the radon quadrupole relaxation rate $\Gamma_2$.  The spin-exchange rate is dependent on the spin-exchange cross section $\sigma_{SE}$ and the cell temperature $T$ through the equation
\begin{equation}
\gamma_{SE}=\left[\mathrm{Rb}\right] \langle\sigma_{SE} v\rangle,
\end{equation}
in which the temperature dependence enters through [Rb] and the velocity-averaged rate constant $\langle\sigma_{SE} v\rangle$.  Quadrupole relaxation ($\Gamma_2$) has been shown to be much stronger than dipole relaxation ($\Gamma_1$) in $^{131}$Xe and $^{21}$Ne \cite{Wu90,Chup90}.
 Thus, $\Gamma_1$ is set to zero in the model.  Wall relaxation is modeled by an Arrhenius-type temperature dependence \cite{Happ72}, i.e.
\begin{equation}
\Gamma_2(T)=\Gamma_2^{\infty}e^{T_0/T},
\end{equation}
where $k_B T_0$ is the binding energy of the radon to the cell wall.  The set of rate equations for the populations of the six nuclear sublevels of $^{209}$Rn are written in terms of the rubidium polarization, $\gamma_{SE}$, and $\Gamma_2$, and an analytic equation for the steady-state sublevel populations is obtained.  Selection rules are used to calculate the sublevel populations of the $^{209}$At populated by electron capture.  The angular distribution of the $^{209}$At gamma rays depends on the moments of the populations, the spins of the states, and the multipolarity of the transitions \cite{Tolh53}.  Table \ref{tab:gammas} includes the multipole mixing ratios ($\delta$) of the four prominent lines and the spin transitions involved.  Following the convention of Ref. \cite{Hart54}, $\delta^2$=a$_1^2/$a$_2^2$, where a$_1=1$ in a pure dipole transition, a$_2=1$ in a pure quadrupole transition, and a$_1^2+$a$_2^2=1$.  The angular distributions are expressed as a function of $\sigma_{SE}$, $\Gamma_2^{\infty}$, $T_0$, and $\delta$ for each cell temperature under consideration.

\begin{figure}[t]
\begin{center}
\includegraphics[width=87mm]{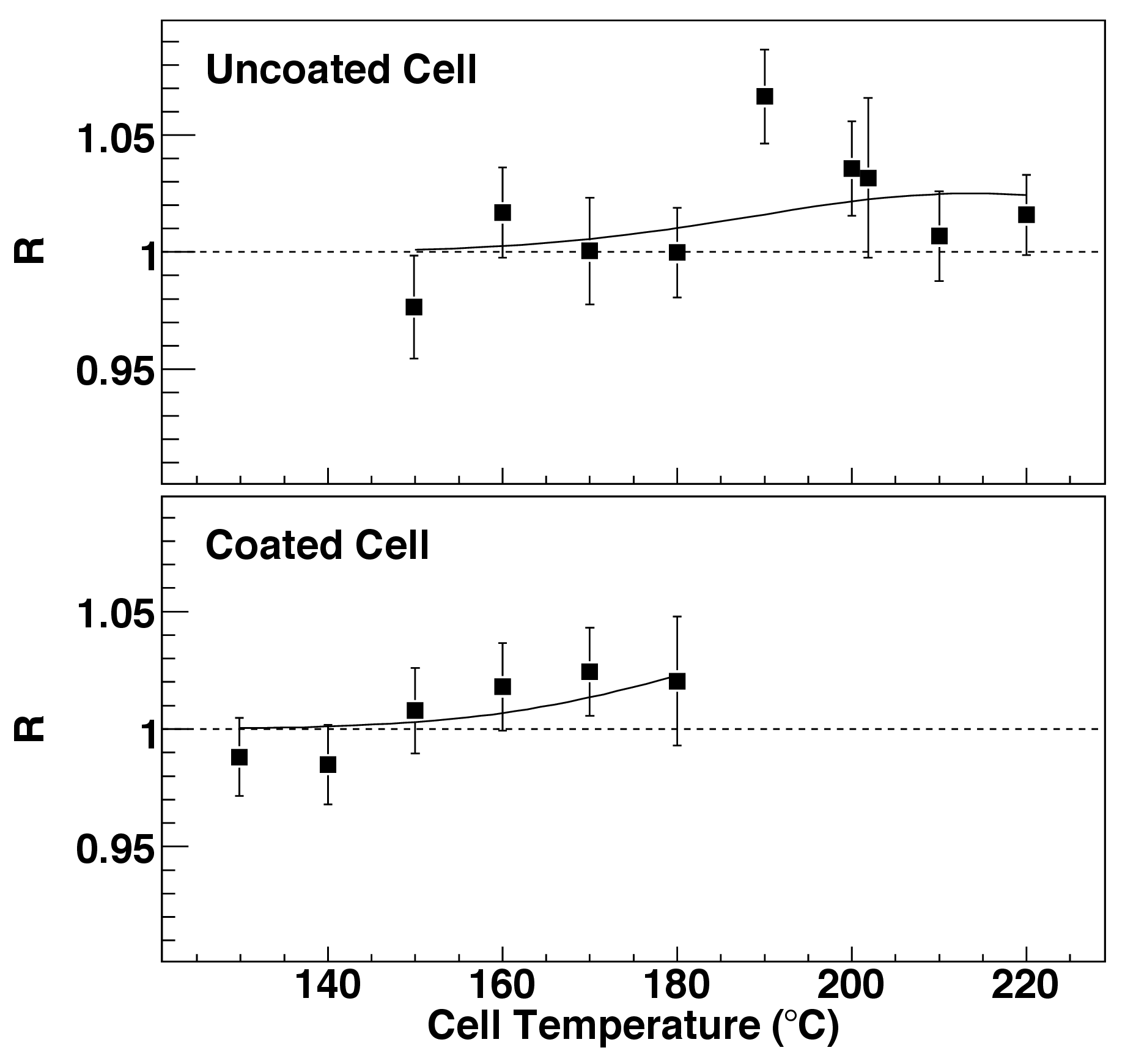}
\end{center}
\caption{Anisotropy data for the 745 keV $^{209}$Rn gamma ray displayed in the same manner as the 337 keV data in Fig. \ref{fig:fig3}.  The solid curves are fits to the mixing ratio $\delta$ given the value of $\Gamma_2^\infty$ obtained from the 337 keV data in the corresponding cell type, as described in the text.}
\label{fig:fig4}
\end{figure}

   The 337, 408, and 745 keV lines were studied in detail. The 689 keV line could not be reliably extracted from the spectra due to the proximity of a gamma-ray line resulting from $^{209}$At decay.  The anisotropy for each gamma ray is given by the ratio
\begin{equation}
R = \frac{(n_{0^\circ}/n_{90^\circ})_{LON}}{(n_{0^\circ}/n_{90^\circ})_{LOFF}},
\end{equation}
where $n_{0^\circ}/n_{90^\circ}$ is the background-corrected ratio of counts in the photopeak at the 0$^\circ$ detector to counts in the 90$^\circ$ detector calculated for data with the laser on (LON) and with the laser off (LOFF).  The 337 and 745 keV lines show anisotropies that are consistent with those measured in Ref. \cite{Kita88}.  The uncoated-cell 408 keV line (not shown here) shows no statistically significant anisotropy, also consistent with Ref. \cite{Kita88}.  
  
The solid lines in the figures were generated using the optical pumping/spin-exchange model.  In Fig. \ref{fig:fig3}, data for the 337 keV line are shown.  Since $\delta$ is known for this transition, the data can be used to extract the relaxation rate $\Gamma_2^\infty$, assuming values for $\sigma_{SE}$ and $T_0$.  For $\sigma_{SE}$, the estimate calculated by Walker \cite{Walk89} ($2.5\times10^{-5} \rm{\AA}^2$) was used.  An estimate for $T_0$ (400) was made based on the $^{131}$Xe measurements reported in Ref. \cite{Wu90}.  For the 337 keV data, best-fit values for $\Gamma_2^{\infty}$ of $0.05\pm0.01$ Hz for the uncoated cell and $0.032\pm0.009$ Hz for the coated cell were found.  These fits are displayed as the solid lines in Fig. \ref{fig:fig3}.  There are multiple ways to generate temperature-dependent angular distributions that agree with the data using different values of $\sigma_{SE}$, $T_0$, and $\Gamma_2^\infty$, so the current data do not allow best fit values for all three parameters simultaneously.
   
   With the polarization and relaxation parameters fixed by the 337 keV data, the 745 keV and 408 keV data can be fit for $\delta$.  The uncoated-cell 745 keV data allow (95\% C.L.) values of $\delta$ in the ranges $\delta<0.19$ and $\delta>10.9$.  Previous measurements \cite{ToI8} exclude the range $\delta<2.86$, and this work significantly improves that constraint.  A similar analysis for the uncoated-cell 408 keV data allows the range $0.01<\delta<0.68$ (95\% C.L.).  This establishes an upper limit of $\delta<0.68$.  The coated-cell data provide much weaker constraints.  These results are not very sensitive to the values of $\sigma_{SE}$, $T_0$, or to knowledge of the cell temperature and laser power.  For example, if the temperature were ten degrees lower than indicated, the limit for the 745 keV transition would become $\delta>11.1$.  With a change of temperature in the model, the best-fit value of $\Gamma_2^\infty$ changes significantly to offset changes in the temperature-dependent values of  the rubidium polarization and the spin-exchange rate.  The best fit ($\delta=\infty$) is plotted for the 745 keV line in Fig. \ref{fig:fig4}.
   
\begin{table}[bp]
\vspace{-.1cm}
\caption{The four main $^{209}$Rn gamma ray lines \cite{ToI8} (from the Table of Isotopes) are listed together with the corresponding absolute intensity I$_\gamma$, spin transition, and multipole mixing ratio ($\delta$).}
\vspace{.1cm}
\label{tab:gammas}
\begin{ruledtabular}
\begin{tabular}{lccccc|}
E$_\gamma$ & I$_\gamma$ & Spin &$\delta$ & $\delta$\\
(keV) & (\%) & Transition & (Ref. \cite{ToI8}) & (this work) \\
\hline
337.45 & 14.5 & 7/2$\rightarrow$7/2 & $\infty$ & - \\
408.32 & 50.3 & 7/2$\rightarrow$9/2 & 0 & $<$0.68  \\
689.26 & 9.7 & 7/2$\rightarrow$7/2 & $>$3.57 & -  \\
745.78 & 22.8 & 7/2$\rightarrow$9/2 & $>$2.86 & $>$10.9 \\
\end{tabular}
\end{ruledtabular}
\end{table}

In summary, radon has been collected and polarized in a glass cell, the temperature dependence of anisotropies has been measured, and new constraints have been set on the multipole mixing ratios for $^{209}$At transitions.  These results are significant for an EDM measurement being planned at TRIUMF.  The ISAC facility is expected to produce $^{223}$Rn at a rate of $10^7/\textrm{s}$, more than 1000 times the rates for this work.  At this rate, after two half-lives of implantation in a zirconium foil, approximately 500 million radon atoms will have been collected.  The precession frequency of polarized radon in an electric field is proportional to the EDM, $d$.  The expected uncertainty in $d$, when measured using the anisotropy signal, is
\begin{equation}
\delta_d=\frac{\hbar}{2 \textrm{E} T_2}\sqrt{\frac{1}{A^2(1-B)^2N}}  ,
\end{equation}
where E is the magnitude of the electric field, $T_2$ is the coherence time of the polarization,  $A$ is the analyzing power of the measurement (i.e., the change in the signal due to a change in polarization), $N$ is the total number of photons detected, and $B$ is the fraction of those photons due to background.  Using $T_2=$ 30 s, based on the uncoated-cell relaxation time, $A=0.1$ and $B=0.01$, which are consistent with this work, and $\textrm{E}=$ 5 kV/cm \cite{Rose01}, a sensitivity of better than $10^{-26}$$e\cdot \textrm{cm}$ is possible with $N=10^{12}$ gammas. Assuming a factor of 400 greater sensitivity for $^{223}$Rn, this would improve sensitivity to sources of CP violation by an order of magnitude or more compared to the limits set by the $^{199}$Hg measurements.

\begin{acknowledgments}
This work was supported by the U. S. National Science Foundation, the Department of Energy, and the Natural Sciences and Engineering Research Council of Canada.
\end{acknowledgments}

\end{document}